\documentclass{elsart1p}

 \usepackage{graphicx}

\begin{document}

\begin{frontmatter}



\title{Hyperon-Quark Mixed Phase in Compact Stars}


\author[jaea]{Toshiki Maruyama},
\author[jaea]{Satoshi Chiba},
\author[infn]{Hans-Josef Schulze},
\author[kyoto]{Toshitaka Tatsumi}

\address[jaea]{Japan Atomic Energy Agency, Tokai, Ibaraki 319-1195, Japan}
\address[infn]{INFN Sezione di Catania, Via Santa Sofia 64, I-95123 Catania, Italy}
\address[kyoto]{Department of Physics, Kyoto University, Kyoto 606-8502, Japan}

\begin{abstract}
We investigate the properties of the hadron-quark mixed phase in compact stars
using a Brueckner-Hartree-Fock framework for hadronic matter and the MIT bag model for quark matter.
We find that the equation of state of the mixed phase is similar to that given
by the Maxwell construction. 
The composition of the mixed phase, however, is very different
from that of the Maxwell construction;
in particular, hyperons are completely suppressed.
\end{abstract}

\end{frontmatter}

\def\tsurf{\sigma}
\def\vc{V_{\rm C}}
\def\rv{{\bf r}}

\section{Introduction}

It is well known that hyperons appear at several times 
normal nuclear density and lead to a strong softening of the equation of state (EOS)
with a consequent substantial reduction of the maximum neutron star mass. 
Actually the microscopic Brueckner-Hartree-Fock approach
gives much lower masses than current observation values of $\sim 1.5M_\odot$. 

On the other hand, the hadron-quark deconfinement transition is believed 
to occur in hot and/or high-density matter.
Taking EOS of quark matter within the MIT bag model, 
the maximum mass can increase to the Chandrasekhar limit once 
the deconfinement transition occurs in hyperon matter \cite{hypns,bal}.
%
Since the deconfinement transition from hadron to quark phase may 
occur as a first-order phase transition, 
the hadron-quark mixed phase should appear, where charge density as well as baryon number density 
is no more uniform. 
Owing to the interplay of the Coulomb interaction 
and the surface tension, the mixed phase can have exotic
shapes called pasta structures \cite{mar}.

The bulk Gibbs calculation of the mixed phase, 
without the effects of the Coulomb interaction and surface tension, 
leads to a broad region of the mixed phase (MP) \cite{gle92}. 
However, if one takes into account the geometrical structures in the
mixed phase and applies the Gibbs conditions, one may find that MP is
considerably limited and thereby EOS approaches to the one given by 
the Maxwell construction (MC)  
\cite{mar}.

In this report we explore the EOS and the structure of the mixed phase during the
hyperon-quark transition, 
properly taking account of
the Gibbs conditions.

\section{Numerical Calculation}

The numerical procedure to determine the EOS and the
geometrical structure of the MP is similar to that 
explained in detail in Ref.~\cite{mar}.
We employ a Wigner-Seitz approximation in which
the whole space is divided into equivalent Wigner-Seitz 
cells with a given geometrical symmetry,
sphere for three dimension (3D), cylinder for 2D, and slab for 1D.
A lump portion made of one phase is embedded in the other phase and thus 
the quark and hadron phases are separated in each cell.
A sharp boundary is assumed between the two phases and the surface energy
is taken into account in terms of a surface-tension parameter $\tsurf$.
The energy density of the mixed phase is thus written as
%
$
 \epsilon = {1\over {V_W}} \left[ 
 {\int}_{V_H} d^3 r \epsilon_H({\rv})+
 {\int}_{V_Q} d^3 r \epsilon_Q({\rv})+
 {\int}_{V_W} d^3 r \left( \epsilon_e({\rv}) + {(\nabla \vc({\rv}))^2\over 8\pi e^2} \right)
 + \tsurf S \right] \:,
%
$
where the volume of the Wigner-Seitz cell $V_W$ is the sum of 
those of hadron and quark phases $V_H$ and $V_Q$,
$S$ the quark-hadron interface area.
$\epsilon_H$, $\epsilon_Q$ and $\epsilon_e$ 
are energy densities of hadrons, quarks and electrons,
which are 
$\rv$-dependent since they are functions of
local densities $\rho_a(\rv)$ ($a=n,p,\Lambda,\Sigma^-,u,d,s,e$). 
The Coulomb potential $\vc$ is obtained by solving the Poisson equation.
For a given density $\rho_B$, the optimum dimensionality of the cell,
the cell size $R_W$, the lump size $R$,
and the density profile of each component
are searched for to give the minimum energy density.
%
We employ $\tsurf=40\;\rm MeV\!/fm^2$ 
in the present study. 

To calculate $\epsilon_H$ in the hadron phase,
we use the Thomas-Fermi approximation for the kinetic energy density.
The potential-energy density is 
calculated by the nonrelativistic BHF approach \cite{hypns}
based on microscopic
NN and NY potentials that are fitted to scattering phase shifts.
Nucleonic three-body forces are included in order to (slightly) shift
the saturation point of purely nucleonic matter to the empirical value.

For the quark phase,
we use the MIT bag model with 
massless $u$ and $d$ quarks and massive $s$ quark with $m_s= 150$ MeV.
The energy density $\epsilon_Q$ consists of the kinetic term by the Thomas Fermi approximation,
the leading-order one-gluon-exchange term \cite{jaf}
proportional to the QCD fine structure constant $\alpha_s$, 
and the bag constant $B$.
We here use $B=100$ $\rm MeV/fm^3$ and
$\alpha_s=0$ to get the quark EOS which crosses the hadronic one
at an appropriate baryon density.

\section{Hadron-Quark Mixed Phase}

\begin{figure*}
\begin{minipage}{0.48\textwidth}
\includegraphics[width=0.90\textwidth]{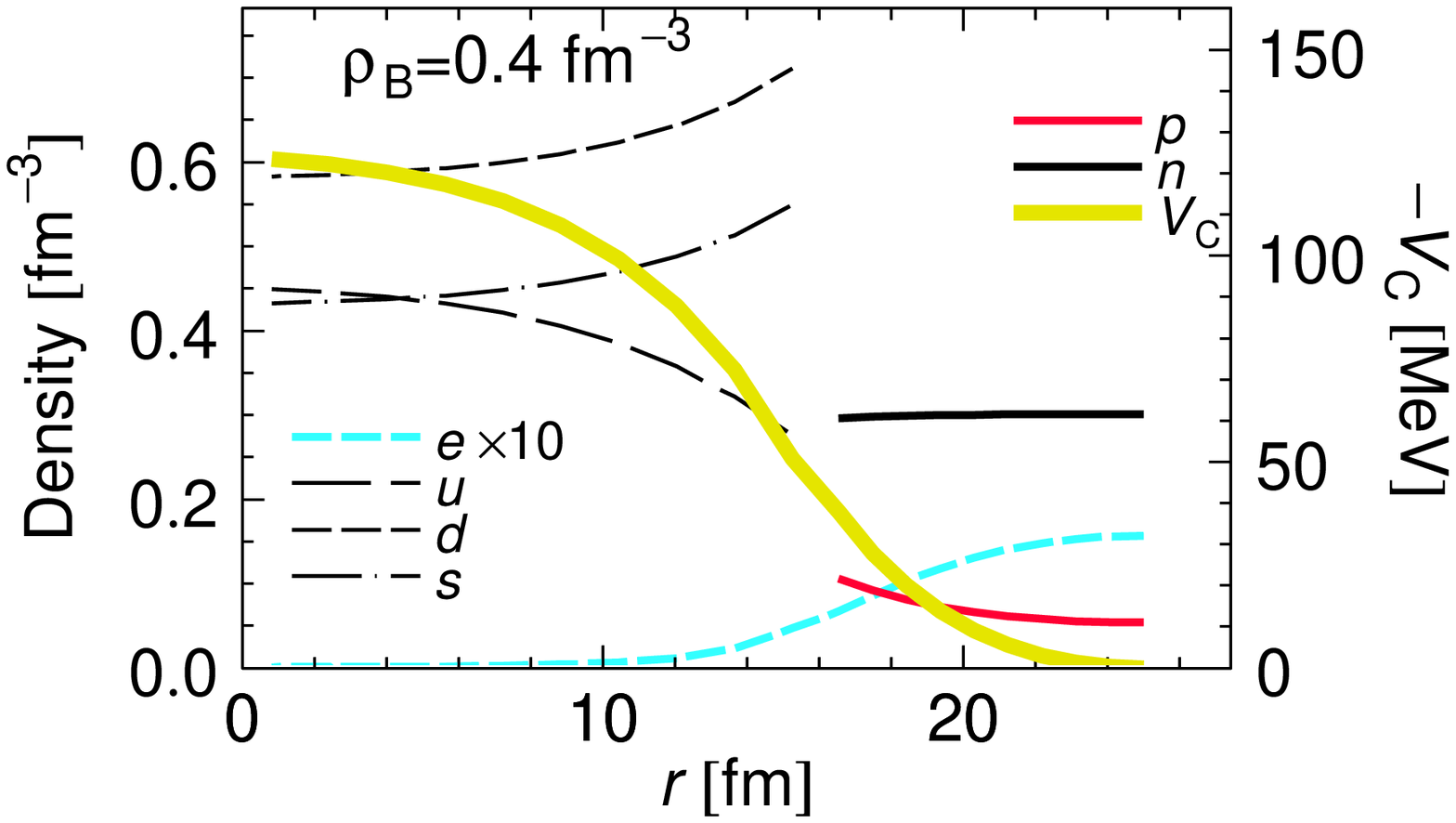}
\caption{
Density profiles 
and Coulomb potential $\vc$ 
within a 3D (quark droplet) Wigner-Seitz cell
of the MP at $\rho_B=0.4$ fm$^{-3}$.
The cell radius and the droplet radius are $R_W=26.7$ fm
and $R=17.3$ fm, respectively.
}
\label{figProf}
\end{minipage}
%
\hspace{\fill}
\begin{minipage}{0.48\textwidth}
\includegraphics[width=0.85\textwidth]{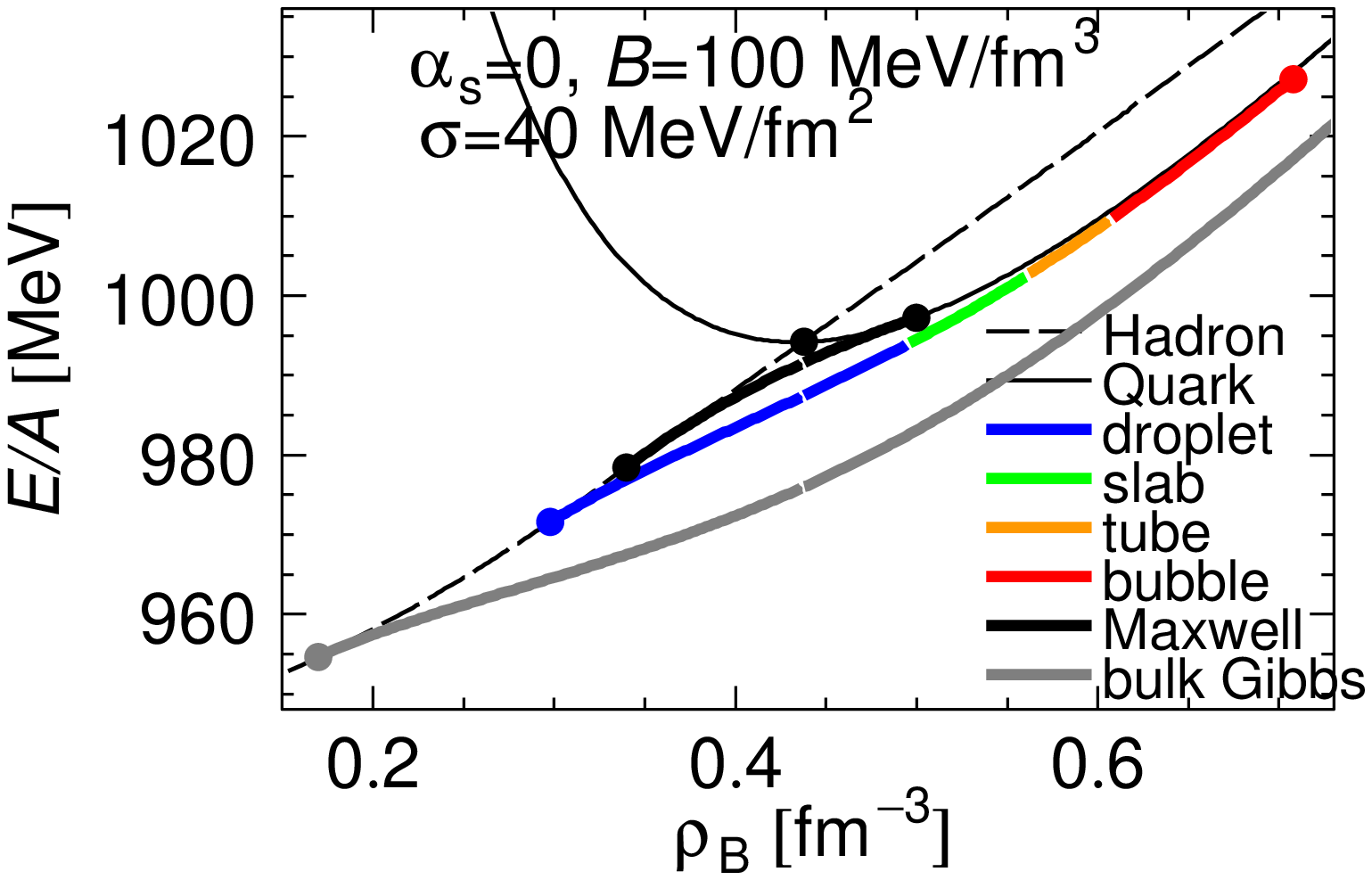}
\caption{
EOS of the MP (thick curves)
in comparison with pure hadron and quark phases (thin curves).
hadron ($\rho_B<0.44$ fm$^{-3}$)
or quark ($\rho_B>0.44$ fm$^{-3}$) phases.
Each segment of the MP is chosen by minimizing the energy.
}
\label{figEOS}
\end{minipage}
\end{figure*}

Figure~\ref{figProf} illustrates an example of 
the density profile in a 3D cell. 
One can see the non-uniform density distribution of each particle species
together with the finite Coulomb potential; charged particle
distributions are rearranged to screen the Coulomb potential.
The quark phase is negatively charged, so that 
$d$ and $s$ quarks are repelled to the phase boundary, 
while $u$ quarks gather at the center.
The protons in the hadron phase are attracted by the negatively charged 
quark phase, while the electrons are repelled.

Figure~\ref{figEOS} compares the resulting EOS 
with that of the pure hadron and quark phases.
The thick black curve indicates the case of the MC,
while the colored line indicates the MP
in its various geometric realizations
starting 
with a quark droplet structure
and ending 
with a bubble structure.
Note that the charge screening effect, combined with the surface tension, 
makes the non-uniform structures mechanically less stable
and limits the density region of the MP \cite{mar}.
Consequently the energy of the MP is only slightly lower than that of the MC. 
However, the structure and the composition of the MP
are very different from those of the MC, 
\begin{figure}
\includegraphics[width=0.80\textwidth]{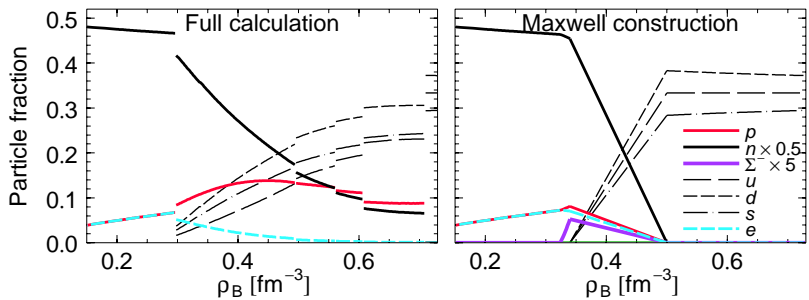}
\caption{
Particle fractions 
in the MP by the full calculation (left panel) and
the MC (right panel).
}
\label{figRatio}
\end{figure}
which is demonstrated in Fig.~\ref{figRatio}, where we compare the 
particle fractions as a function of baryon density in the full calculation (left panel)
and the MC (right panel).
One can see that the compositions are very different in two cases.
In particular, a relevant hyperon ($\Sigma^-$) fraction is only present in the MC.
Non-uniform structure and suppression of hyperons in the mixed
phase should have   
some implications on the neutrino transport and cooling of pulsars.

\end{document}